\documentclass[prd,onecolumn,showpacs,showkeys,floatfix,amsmath,amssymb,floatfix,english]{revtex4}
\usepackage{graphicx}
\usepackage{slashed}
\usepackage{amsmath}
\usepackage{autobreak}
\usepackage[colorlinks,linkcolor=blue,anchorcolor=blue,citecolor=blue]{hyperref}
\usepackage[capitalize]{cleveref}
\allowdisplaybreaks

\newcommand{\qq}[1]{\left\langle \bar{#1} #1 \right\rangle}
\newcommand{\qgq}[1]{\left\langle g_s \bar{#1} \sigma \cdot G #1 \right\rangle}
\renewcommand{\gg}{\left\langle g_s^2 G^2 \right\rangle}

\graphicspath{{figs/}}
\begin{document}
\title{QCD sum rule study for hidden-strange pentaquarks}
\author{Pengfei Yang}
\author{Wei Chen}
\email{chenwei29@mail.sysu.edu.cn}
\affiliation{School of Physics, Sun Yat-Sen University, Guangzhou 510275, China}

\begin{abstract}
Inspired by the LHCb's observations of hidden-charm $P_{c(s)}$ states, we study their hidden-strange analogues $P_s$ states in both $[udu][\bar ss]$ and $[uds][\bar su]$ configurations. We investigate the $P_s$ pentaquark states in $p\eta^\prime$, $p\phi$, $\Lambda K$, $\Sigma K$ and $\Sigma^\ast K^\ast$ structures with $J^P = \frac{1}{2}^-$ and $\Sigma ^\ast K$ and $\Sigma K^\ast$ with $J^P = \frac{3}{2}^-$, and calculate their masses in the framework of QCD sum rules. Our numerical results show that the extracted hadron masses for all the $p\eta^\prime$, $p\phi$, $\Lambda K$, $\Sigma K$ and $\Sigma^\ast K^\ast$ structures are much higher than $\Sigma K$ mass threshold and masses for $\Sigma ^\ast K$ and $\Sigma K^\ast$ also higher than threshold of corresponding hadron, so that no bound state exists in such channels, which is consistent with the experimental status to date.
\end{abstract}

\maketitle

\section{Introduction}
Quantum chromodynamics (QCD) is the fundamental theory of strong interaction. Hadrons are $q\bar q$ mesons and $qqq$ baryons in the conventional quark model~\cite{Gell-Mann:1964ewy,1964-Zweig-p-}. However, QCD itself allows the existence of multiquarks. 
In 2015, the LHCb Collaboration reported two hidden-charm pentaquark states $P_c(4380)$ and $P_c(4450)$ in 
the $J/\psi p$ invariant mass spectrum of $\Lambda_b^0\to J/\psi pK$ decays~\cite{Aaij:2015tga}. The $P_c(4450)$ was then further found to be separated 
into two structures $P_c(4440)$ and $P_c(4457)$ in the same process, in which a new narrow resonance 
$P_c(4312)$ was also discovered~\cite{Aaij:2019vzc}. Recently, the LHCb Collaboration announced the first evidence of a hidden-charm pentaquark state with strangeness, $P_{cs}^0(4459)$, in the $J/\psi\Lambda$ invariant mass distribution of $\Xi_b^0\to J/\psi\Lambda K$ decays~\cite{LHCb:2020jpq}. The observations of these hidden-charm pentaquarks have been 
immediately raised extensively theoretical interests, considering them as $\bar D^{(\ast)}\Sigma_c^{(\ast)}$ ($\bar D^{(\ast)}\Xi_c$) hadron molecules or compact pentaquarks with quark contents 
$\bar ccuud$ ($\bar ccuds$)~\cite{2016-Chen-p1-121,2018-Guo-p15004-15004,2019-Liu-p237-320,2019-Chen-p51501-51501,2021-Chen-p409-409}. 

If the above interpretations for $P_c$'s are correct, the analogous effects could also be expected at the hidden-strange 
$P_s=\bar ssuud$ pentaquark. Actually, the study for existence of hidden-strange pentaquark is much earlier than the observations 
of the $P_c$ states. Experiments and analysis of pentaquark composed of light quark can date back to $N\bar{K}$ molecular state, which were explained by $\Lambda(1405)$ since 1960~\cite{DALITZ1960307,Kaiser:1995eg,Oset:1997it,Krippa:1998us,Oller:2000fj,Jido:2010ag,Hall:2014uca}. But there are almost no results of light pentaquark. In 1994-1999, the experiments with the SPHINX spectrometer reported a resonance structure $X(2000)$ as a candidate of $\bar ssuud$ state in the proton diffractive reactions $p+N(C)\to X(\Sigma^0K^+)+N(C)$~\cite{SPHINX:1994,SPHINX:1995wtz,SPHINX:1994yxk,SPHINX:1999rcq}. Although the resonance extracted had fairly poor statistics, it still inspired some theoretical studies for the existence of the hidden-strange pentaquarks. 
In Ref.~\cite{Williams:2003tj}, R. Williams and P. Gueye studied the color octet-octet $[q\bar s][uds]$ configuration in a non-relativistic quark molecular model and predicted four candidates for $P_s$ states. By using the quark delocalization color screening model (QDCSM), Huang {\it et al.} calculated the effective potential, masses and decay widths for the $\Sigma^{(\ast)}K^{(\ast)}$ molecular states and found that the interactions between $\Sigma^{(\ast)}$ and $K^{(\ast)}$ were strong enough to form bound states~\cite{Huang:2018ehi}. 

Besides, a $\phi-N$ bound state was proposed by Gao {\it et al.} that the QCD van der Waals attractive potential is strong enough to bind a $\phi$  meson onto a nucleon inside a nucleus to form a bound state~\cite{Gao:2000az}. The calculations in the chiral SU(3) quark model~\cite{Huang:2005gw,Liu:2018nse}, lattice QCD~\cite{Beane:2014sda}, chiral soliton model~\cite{Kopeliovich:2015vqa} and QDCSM~\cite{Gao:2017hya} also supported the existence of the $\phi N$ bound states. In Ref.~\cite{Xie:2017mbe}, the authors proposed a possible $\phi p$ resonance in the $\Lambda^+_c\to\pi^0\phi p$ decay by considering a triangle singularity mechanism. In Ref.~\cite{Lebed:2015fpa}, Lebed used a $[su]_{\bf \bar 3}[\bar sud]_{\bf 3}$ diquark-triquark model to investigate the possible existence of $P_s$ pentaquarks, and further proposed to create such states through the decay $\Lambda_c^+\to P_s\pi^0\to\phi p\pi^0$~\cite{Lebed:2015dca}. Unfortunately, the Belle Collaboration searched for the $\Lambda_c^+\to\phi p\pi^0$ decays in 2017 and found no evidence for an intermediate hidden-strange pentaquark decay $P_s\to\phi p$~\cite{Belle:2017tfw}.

Comparing to the hidden-charm $P_c$ states observed in the $J/\psi p$ final state, one possible reason for the absence of $P_s$ pentaquarks may be the limited phase space for $P_s\to\phi p$ decay. Nevertheless, it is still interesting to investigate the properties of $P_s$ states in theory. In this work, we shall study the $\bar ssuud$ systems in both $[udu][\bar ss]$ and $[uds][\bar su]$ configurations with quantum numbers $J^P=\frac{1}{2}^-$ by using the QCD sum rules~\cite{Shifman:1978by,REINDERS19851,Colangelo:2000dp}.

This paper is organized as follows. In Sec.~\ref{QSR}, we construct the interpolating currents for hidden-strange pentaquarks in both $[udu][\bar ss]$ and $[uds][\bar su]$ configurations with $J^P=\frac{1}{2}^-$, and then evaluate the correlation functions and spectral densities for these interpolating currents. In Sec.~\ref{NR}, we perform QCD sum rule analyses to extract the mass spectra for $P_s$ pentaquarks. The last section is a brief summary and discussion.

\section{QCD sum rules for hidden-strange pentaquarks}\label{QSR}
In this section, we introduce the QCD sum rules for the hidden-strange pentaquarks with quark contents $qqqs\bar s$. The start of QCD sum rules is the 
two-point correlator
\begin{eqnarray}
    \Pi(q) &=& i \int{\mathrm{d^4}x e^{i q \cdot x} \left \langle 0 \left | T \left\{ \eta(x)\bar{\eta}(0) \right\} \right |0\right\rangle}\, ,\nonumber\\
    \Pi_{\mu\nu}(q) &=& i \int{\mathrm{d^4}x e^{i q \cdot x} \left \langle 0 \left | T \left\{ \eta_{\mu}(x)\bar{\eta}_{\nu}(0) \right\} \right |0\right\rangle}\, ,
    \label{twopoint}
\end{eqnarray}
where $\eta(x)$ is a hidden-strange pentaquark interpolating current. There are two possible color configurations for such  pentaquark operators: $[\epsilon^{abc}q_aq_bq_c][\bar s_ds_d]$ and $[\epsilon^{abc}q_aq_bs_c][\bar s_dq_d]$, where $a, b, c, d$ are color indices, $q$ denotes a light quark and $s$ is a strange quark. We shall use both these two types of interpolating currents to investigate the hidden-strange pentaquark systems. To construct the pentaquark currents, we use the following operators for the $L=0$ baryons~\cite{1981-Ioffe-p317-341,1983-Ioffe-p67-67,Colangelo:2000dp}
\begin{align}
\nonumber
\eta^p &=\epsilon^{abc}\left[(u^T_a C d_b)\gamma_5 u_c-(u^T_a C \gamma_5 d_b) u_c\right]\, ,
\\ \nonumber \eta_\mu^p &=\epsilon^{abc}\left[(u^T_a C d_b)\gamma_{\mu} u_c-(u^T_a C \gamma_5 d_b)\gamma_{\mu} \gamma_5 u_c\right]\, ,
\\ \eta^\Lambda &=\epsilon^{abc}\left[(u^T_a C d_b)\gamma_5 s_c-(u^T_a C \gamma_5 d_b) s_c\right]\, ,
\\ \nonumber \eta^\Sigma &=\epsilon^{abc}\left[(u^T_a C \gamma_{\mu} d_b)\gamma_{5}\gamma^{\mu}s_c\right]\, ,
\\ \nonumber \eta_\mu^{\Sigma^\ast} &=\epsilon^{abc}\left[2(u^T_a C \gamma_{\mu} s_b)u_c + (u^T_a C \gamma_{\mu} u_b)s_c \right]\, ,
\end{align}
in which $T$ denotes transposition and $C$ is the charge conjugation matrix. Together with the operators $\bar q_a\gamma_5 q_a (0^-)$, $\bar q_a\gamma_\mu q_a (1^-)$ for the $L=0$ mesons, we compose the following hidden-strange pentaquark currents with $J^P = \frac{1}{2}^{-}$ 
\begin{align}
\nonumber
    \eta_1 &= \epsilon^{abc}\left[(u^T_a C d_b)\gamma_5 u_c-(u^T_a C \gamma_5 d_b) u_c\right]\left[\bar{s}_d\gamma_5 s_d\right]\, ,\\ \nonumber
    \eta_2 &= \epsilon^{abc}\left[(u^T_a C d_b)\gamma_{\mu} u_c-(u^T_a C \gamma_5 d_b)\gamma_{\mu} \gamma_5 u_c\right]\left[\bar{s}_d\gamma^{\mu} s_d\right]\, ,\\ \label{pentaquarkcurrents}
    \eta_3 &= \epsilon^{abc}\left[(u^T_a C d_b)\gamma_5 s_c-(u^T_a C \gamma_5 d_b) s_c\right]\left[\bar{s}_d\gamma_5 u_d\right]\, ,\\ \nonumber
    \eta_4 &= \epsilon^{abc}\left[(u^T_a C \gamma_{\mu} d_b)\gamma_{5}\gamma^{\mu}s_c\right]\left[\bar{s}_d\gamma_5 u_d\right]\,,\\ \nonumber
    \eta_5 &=\epsilon^{abc}\left[2(u^T_a C \gamma_{\mu} s_b)u_c + (u^T_a C \gamma_{\mu} u_b)s_c \right] \left[\bar{s}_d\gamma^{\mu}d_d\right]\, , 
\end{align}
which can couple to $p\eta^\prime$, $p\phi$, $\Lambda K$, $\Sigma K$ and $\Sigma^\ast K^\ast$ respectively. And we also compose the following hidden-strange pentaquark currents with $J^P = \frac{3}{2}^-$
\begin{align}
  \eta_{6\mu} &=\epsilon^{abc}\left[(u^T_a C \gamma_{\nu} d_b)\gamma_{5}\gamma^{\nu}s_c\right]\left[\bar{s}_d\gamma_\mu u_d\right]\, ,\\ \nonumber
	\eta_{7\mu} &=\epsilon^{abc}\left[2(u^T_a C \gamma_{\mu} s_b)u_c + (u^T_a C \gamma_{\mu} u_b)s_c \right] \left[\bar{s}_d\gamma_5 d_d\right]\, ,
\end{align}
which can couple to $\Sigma K^\ast$ and $\Sigma^\ast K$ respectively.In principle, these interpolating currents can couple to pentaquark states with both negative and positive parities 
\begin{align}
    \langle 0| \eta(x)| X^- \rangle &= f_{X^-} u(q)\, ,  \label{coupling1}
    \\
    \langle 0| \eta(x)| X^+ \rangle &= if_{X^+} \gamma_5u(q)\, , \label{coupling2}
\end{align}
where $u(q)$ is the Dirac spinor, $f_{X^-}$ and $f_{X^+}$ are the coupling constants. 
Accordingly the two-point correlators induced by these pentaquark currents can be written as 
\begin{align}
  \nonumber
    \Pi^-(q) &= i \int{\mathrm{d^4}x e^{i q \cdot x} \left \langle 0 \left | T \left\{ \eta(x)\bar{\eta}(0) \right\} \right |0\right\rangle} 
    \\ &= \slashed{q}\Pi_q(q^2)+\Pi_m(q^2)\, , \label{pi-}
    \\ \nonumber
    \Pi^+(q) &= i \int{\mathrm{d^4}x e^{i q \cdot x} \left \langle 0 \left | T \left\{ \eta^\prime(x)\bar{\eta^\prime}(0) \right\} \right |0\right\rangle} 
    \\&= \slashed{q}\Pi_q(q^2)-\Pi_m(q^2)\, , \label{pi+}
\end{align}
where $\Pi_q(q^2)$ and $\Pi_m(q^2)$ are the invariant functions proportional to $\slashed q$ and $1$, respectively. One notes that these two invariant functions appear in $\Pi^-(q)$ and $\Pi^+(q)$ at the same time, implying that they contain  hadron information for both the negative and positive parity pentaquark states. However, the signs of $\Pi_m(q^2)$ in $\Pi^-(q)$ and $\Pi^+(q)$ are different due to the coupling definitions in Eqs.~\eqref{coupling1}-\eqref{coupling2}. One can consult Refs.~\cite{Chung:1981cc,Jido:1996ia,Kondo:2005ur,Ohtani:2012ps} for detail discussions.

Besides, $\eta_{6\mu}(x)$ and $\eta_{7\mu}(x)$ can also couple with both $J = \frac{3}{2}$ and $\frac{1}{2}$ channels
\begin{align}
  \Pi_{\mu\nu}(q) = \left(\frac{q_\mu q_\nu}{q^2}-g_{\mu\nu}\right)(\slashed{q}+M_X)\Pi^{3/2}(q)+\cdots\, ,
\end{align}
where $\cdots$ contains the spin $1/2$ components of $\eta_{6\mu}$ or $\eta_{7\mu}$.

At the hadronic level, the invariant functions can be described by the dispersion relation
\begin{equation}
\Pi\left(q^{2}\right)=\frac{\left(q^{2}\right)^{N}}{\pi} \int_{4m_{s}^{2}}^{\infty} \frac{\operatorname{Im} \Pi(s)}{s^{N}\left(s-q^{2}-i \epsilon\right)} d s+\sum_{n=0}^{N-1} b_{n}\left(q^{2}\right)^{n}\, ,
\label{Cor-Spe}
\end{equation}
where $b_n$ are $N$ unknown subtraction constants and will be removed after performing Borel transform. The imaginary part of the correlation function is usually defined as the spectral function 
\begin{align}
\nonumber
\rho (s)\equiv\frac{1}{\pi} \text{Im}\Pi(s)&=\sum_n\delta\left(s-m_n^2\right)\langle 0|\eta|n\rangle\langle n|\bar\eta|0\rangle
\\ &=f_{X}^{2}\delta(s-m_{X}^{2})+\cdots\, , \label{spectral}
\end{align}
in which the “pole plus continuum” parametrization assumption is adopted and ``$\cdots$'' contain contributions from the QCD continuum and higher excited states. The intermediate state $|n\rangle$ can be either positive or negative parity pentaquark.  The parameters $f_{X}$ and $m_{X}$ are the coupling constant and hadron mass of the lowest-lying state. 

The two-point correlation function and spectral function can be evaluated as the functions of various QCD condensates via operator product expansion (OPE) at the quark-gluonic level. We shall use the coordinate space expressions for the light quark and strange quark propagators~\cite{Yang:1993bp}
\begin{align}  
\begin{autobreak}          
i S^{ab}(x) =       
\langle T\{q^a(x)\bar{q}^b(0)\}\rangle\notag=       
\frac{i\delta^{ab}}{2\pi^2x^4}      
+\frac{i}{32\pi^2}t^n_{ab} g_s G^n_{\mu\nu}\frac{1}{x^2}(\sigma^{\mu\nu}\slashed{x}      
+\slashed{x}\sigma^{\mu\nu})      
-\frac{\delta^{ab}}{12}\langle\bar{q}q\rangle\notag      
+\frac{\delta^{ab}x^2}{192}\langle g_s \bar{q}\sigma Gq\rangle      
-\frac{i g_s^2 \langle \bar    {q}q\rangle^2x^2}{2^5\times 3^5}\delta^{ab}\slashed{x}      
- \frac{g_s^2\langle\bar{q}q\rangle\langle G^2\rangle x^4}{2^9\times3^3}\delta^{ab}\notag      
-\frac{m_q\delta^{ab}}{4\pi^2x^2}      
+\frac{m_q}{16\pi^2}t^n_{ab}g_s G^n_{\mu\nu}\sigma^{\mu\nu}\log(-x^2)      
-\frac{\delta^{ab}\langle g_s^2G^2\rangle}{2^9\times3\pi^2}m_q x^2\log(-x^2)\notag      
+\frac{i\delta^{ab}m_q\langle\bar{q}q\rangle}{48}\slashed{x}      
-\frac{i m_q\langle g_s\bar{q}\sigma G q\rangle\delta^{ab}x^2}{2^7\times3^2}\slashed{x}       
- \frac{g_s^2 m_q\langle \bar{q}q\rangle^2}{2^7\times3^5}x^4\delta^{ab}      
+\cdots          
\end{autobreak}
\end{align}
where $q=u, d, s$ quark, $\slashed{x}=x^{\mu}\gamma_{\mu}$ and $t^n_{ab}=\lambda^n_{ab}/2$. In this work, we  evaluate the correlation functions and spectral functions up to dimension-11 condensates. As an example, we show the spectral function for the interpolating current $\eta_1(x)$ as 
\begin{align}
    \begin{autobreak}
        \rho_1^q(s)=
        \frac{s^5}{45875200 \pi ^7}
        -\frac{s^3 m_s \qq{s} }{40960 \pi ^5}
        +\frac{s^3 \gg }{1310720 \pi ^7}
        +\frac{g_s^2 s^2 \qq{u} ^2}{110592 \pi ^5}
        +\frac{g_s^2 s^2 \qq{s} ^2}{110592 \pi ^5}
        +\frac{g_s^2 s^2 \qq{d} ^2}{221184 \pi ^5}
        +\frac{s^2 \qq{s} ^2}{2048 \pi ^3}
        -\frac{415 s^2 m_s \qgq{s} }{1572864 \pi ^5}
        +\frac{127 g_s^3 \qq{d} ^2 \gg }{31850496 \pi ^5}
        -\frac{g_s^2 m_s \qq{d} ^2 \qq{s} }{6912 \pi ^3}
        -\frac{65 m_s \gg  \qgq{s} }{2359296 \pi ^5}
        -\frac{161 s m_s \qq{s}  \gg }{2359296 \pi ^5}
        +\frac{161 \qq{s} ^2 \gg }{589824 \pi ^3}
        +\frac{65 g_s^3 \qq{s} ^2 \gg }{31850496 \pi ^5}
        +\frac{5 g_s^3 \qq{u} ^2 \gg }{663552 \pi ^5}
        +\frac{s \qq{s}  \qgq{s} }{512 \pi ^3}
        +\frac{3 \qgq{s} ^2}{4096 \pi ^3}
        -\frac{g_s^2 m_s \qq{s}  \qq{u} ^2}{3456 \pi ^3}
        +\frac{g_s^2 m_s \qq{s} ^3}{3456 \pi ^3}\, , 
    \end{autobreak}\label{rhoq}
\end{align}
\begin{align}
    \begin{autobreak}
        \rho_1^m(s) = 
        \frac{s^4 \qq{d} }{491520 \pi ^5}
        +\frac{s^2 \qq{d}  \gg }{73728 \pi ^5}
        -\frac{s^2 m_s \qq{d}  \qq{s} }{1536 \pi ^3}
        +\frac{s \gg  \qgq{d} }{65536 \pi ^5}
        +\frac{g_s^2 s \qq{d}  \qq{u} ^2}{10368 \pi ^3}
        +\frac{g_s^2 s \qq{d}  \qq{s} ^2}{10368 \pi ^3}
        -\frac{s m_s \qq{d}  \qgq{s} }{768 \pi ^3}
        -\frac{m_s \qgq{d}  \qgq{s} }{1024 \pi ^3}
        +\frac{\qq{d}  \qq{s}  \qgq{s} }{192 \pi }
        +\frac{\qq{u} ^2 \qgq{d} }{192 \pi }
        +\frac{\qq{d}  \qq{u}  \qgq{u} }{96 \pi }
        +\frac{\qq{d}  \gg ^2}{1179648 \pi ^5}
        +\frac{g_s^2 \qq{u} ^2 \qgq{d} }{41472 \pi ^3}
        +\frac{s \qq{d}  \qq{u} ^2}{96 \pi }
        +\frac{s \qq{d}  \qq{s} ^2}{192 \pi }\, , 
    \end{autobreak}\label{rhom}
\end{align}
where $\rho^q(s)=\text{Im}\Pi_q(s)/\pi$, $\rho^m(s)=\text{Im}\Pi_m(s)/\pi$ are the spectral functions for the invariant structures $\Pi_q(q^2)$ and $\Pi_m(q^2)$ in Eqs.~\eqref{pi-}-\eqref{pi+} respectively. Their dimensions are different 
because $\Pi_q(q^2)$ is proportional to $\slashed{q}$ while $\Pi_m(q^2)$ proportional to 1. As shown in Eqs.~\eqref{rhoq}-\eqref{rhom}, there are more terms in $\rho^q(s)$ than those in $\rho^m(s)$, including the perturbative term. We shall use $\rho^q(s)$ to perform our numerical analyses in the following section. We show the expressions of spectral functions for other interpolating currents in Appendix.~\ref{ros}. One should note that all the correlation functions in this work are evaluated at the leading order of $\alpha_s$. However, it is known that the $\alpha_s$ corrections of the perturbative terms gave important contributions  for the heavy quarkonium systems~\cite{REINDERS19851}. However, we will not consider the $\alpha_s$ corrections here because of the complexity and difficulty of the calculations. Nevertheless, the NLO effects in multiquark systems are still deserved to be studied and we will try to conduct such research in our future work.

The QCD sum rules can be established by assuming that the two-point correlation functions obtained from the hadronic level and quark-gluonic level are equal to each other. After applying the Borel transform, the correlation function can be written as
\begin{equation}
\Pi(M_B^2)\equiv\mathcal{B}_{M_B^2}\Pi(q^2)=\int^{\infty}_{4m_s^2}ds\,e^{-s/M_B^2}\rho(s)\, ,
\end{equation}
in which $M_B$ is the Borel parameter introduced by the Borel transform. It is clear that the Borel transform shall suppress the contributions from the continuum and higher excited states in Eq.~\eqref{spectral}. The lowest-lying resonance is then picked out as. 
\begin{equation}\label{EqL}
f^2_Xe^{-m^2_X/M_B^2}=\int^{s_0}_{4m_s^2}ds\,e^{-s/M_B^2}\rho(s)\, ,
\end{equation}
where $s_0$ is the continuum threshold. The mass of the lowest-lying hadron can be thus extracted as
\begin{equation}\label{mass}
m^2_X(s_0,M_B^2)=\frac{\int^{s_0}_{4m_s^2}ds\,e^{-s/M_B^2}s\rho(s)}{\int^{s_0}_{4m_s^2}ds\,e^{-s/M_B^2}\rho(s)}.
\end{equation}
which is the function of two parameters $M_B^2$ and $s_0$. We shall discuss the detail to obtain suitable parameter working regions in QCD sum rule analysis in next section.

\section{Numerical results}\label{NR}
To perform numerical analyses, we adopt the parameter values for various QCD condensates and quark masses as~\cite{Yang:1993bp,ParticleDataGroup:2020ssz,Narison:2002woh,Gimenez:2005nt}: 
$\langle \bar{u} u\rangle=\langle \bar{d} d\rangle=\langle \bar{q} q\rangle = - (0.24\pm0.01)^3$ GeV$^3$,
 $\langle \bar{q} g_s\sigma G q\rangle  = m_0^2\langle \bar{q} q\rangle$,  $m_0^2=-0.8$ GeV$^2$,
 $\langle \bar{s} s\rangle = (0.8\pm0.1)\langle \bar{q} q\rangle$, 
 $\langle \bar{s} g_s\sigma G s\rangle  = m_0^2\langle \bar{s} s\rangle$,
 $\langle g_s^2GG\rangle = (0.44\pm0.02)$ GeV$^4$, and $\overline{MS} $ strange quark mass $m_s = (0.095\pm0.005)$ GeV. We let $m_u=m_d=m_q=0$, $\langle \bar{u} u\rangle=\langle \bar{d} d\rangle=\langle \bar{q} q\rangle$ and $\langle \bar{u} g_s\sigma\cdot G u\rangle=\langle \bar{d} g_s\sigma\cdot G d\rangle=\langle \bar{q} g_s\sigma\cdot G q\rangle$ for the up and down quarks in the chiral limit. The definition of the coupling constant $g_s$ has a minus sign difference compared to that in Ref.~\cite{REINDERS19851}.

As shown in Eq.~\eqref{mass}, the extracted hadron mass is a function of the Borel mass $M_B$ and continuum threshold $s_0$. To choose suitable working windows for these two parameters, one needs to study the OPE convergence and pole contribution for the correlation function. On the one hand, the perturbative term should be two times larger than nonperturbative contribution to ensure good OPE convergence, which shall limit the lower bound on the Borel parameter $M_B^2$. On the other hand, we require the following pole contribution be larger than $10\%$ to give the upper bound on $M_B^2$  
 \begin{equation}\label{PC}
PC(s_0,M_B^2)=\frac{\int^{s_0}_{4m_s^2}ds\,e^{-s/M_B^2}\rho(s)}{\int^{\infty}_{4m_s^2}ds\,e^{-s/M_B^2}\rho(s)}\, ,
\end{equation}
which is also the function of $M_B^2$ and $s_0$. A proper value of $s_0$ is needed before determining the Borel window. 

\begin{figure}
    \centering
    \includegraphics[scale=0.5]{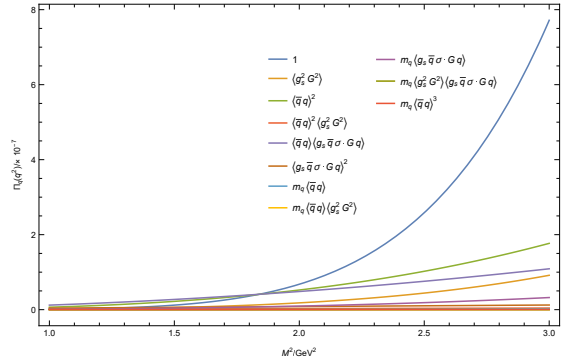}
    \caption{The contributions of each term in the OPE series for the interpolating current $\eta_4(x)$ with $J^P=\frac{1}{2}^-$.}
    \label{fig:lower}    
\end{figure}

\begin{figure}
    \centering
    \includegraphics[scale=0.5]{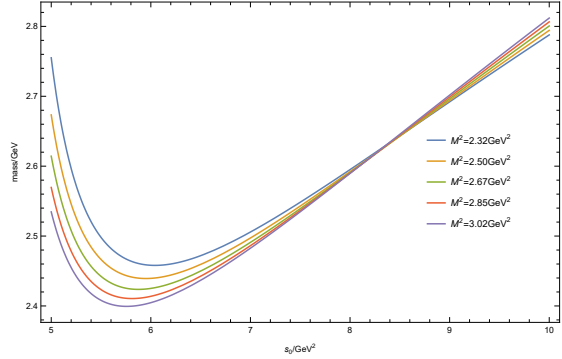}
    \caption{Hadron mass with respect to the threshold parameter $s_0$ with different value of Borel mass for $\eta_4(x)$.}
    \label{fig:s0}    
\end{figure}

We take the interpolating current $\eta_4(x)$ as an example to show our numerical analyses for the hidden-strange $\Sigma K$ pentaquark state. In Fig.~\ref{fig:lower}, we plot the OPE behavior term by term with respect to $M_B^2$. It is shown that the most important nonperturbative contribution is from the $\langle \bar{q}q\rangle^2 $ term. Requiring the perturbative term be two times larger than $\langle \bar qq\rangle^2$ term, one finds the lower bound on the Borel parameter $M_{min}^2 = 2.32$ GeV$^2$.
To choose an optimum value of $s_0$, we show the variation of hadron mass with respect to $s_0$ with different value of Borel mass in Fig.~\ref{fig:s0}. The best choice of the threshold parameter is then fixed as $s_0 = 8.35$ GeV$^2$, around which the hadron mass is very stable against $M_B^2$.

\begin{figure}
    \centering
    \includegraphics[scale=0.5]{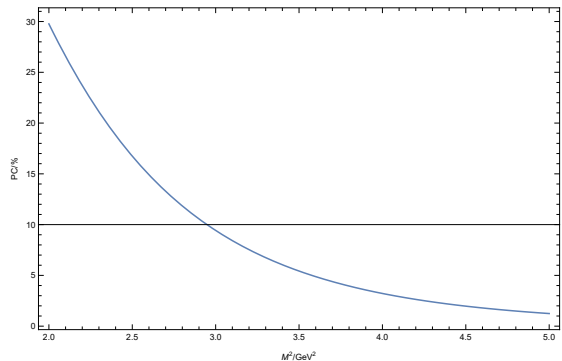}
    \caption{Pole contribution behavior for $\eta_4(x)$ with $s_0 = 8.35$ GeV$^2$.}
    \label{fig:pole}    
\end{figure}

\begin{figure}
    \centering
    \includegraphics[scale=0.6]{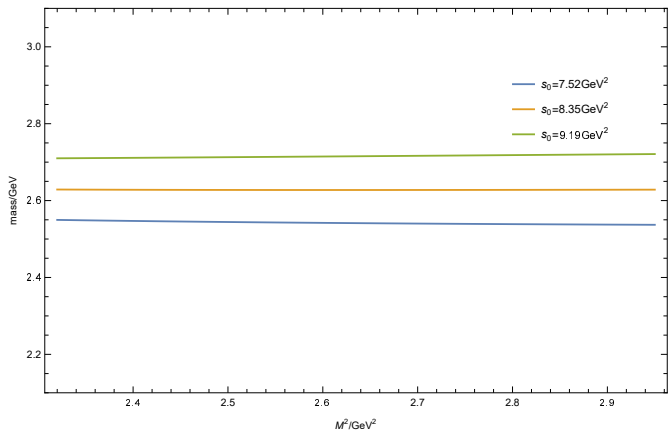}
    \caption{Variations of hadron mass $m_X$ with respect to $M_B^2$ for $\eta_4(x)$.}
    \label{fig:mass}    
\end{figure}

To determine the upper limit of $M_B^2$, we study the pole contribution for $s_0 = 8.35$ GeV$^2$ in Fig.~\ref{fig:pole}. 
We find that the pole contribution is larger than $10\%$ for $M_B^2\leq2.95$ GeV$^2$, which is the upper bound on the Borel window. Finally, we can study the behavior of the Borel curves in Fig.~\ref{fig:mass} by using the parameters $s_0 = 8.35$ GeV$^2$ and Borel window $2.32$ GeV$^2$$\leq M_B^2\leq2.95$ GeV$^2$. One finds that the mass curves are very stable with respect to $M_B^2$ in the these parameter regions. 
The hadron mass for such a hidden-strange $\Sigma K$ pentaquark with $J^P=\frac{1}{2}^-$ is extracted as 
\begin{equation}
m_X=(2.63\pm 0.14\pm 0.13)~\text{GeV}\, ,
\end{equation}
in which the first error comes from the uncertainties of various input QCD parameters (mainly from the quark condensate), while the second error comes from  the uncertainty of the threshold value $s_0$. 
For all other interpolating currents in Eq.~\eqref{pentaquarkcurrents}, we perform similar analyses as the above procedure and obtain the hadron masses in Table~\ref{tablemass}. We find that the mass for the $p\phi$ pentaquark is several MeV below the $M_{p\phi}$ threshold. However, it is still much higher than the mass threshold of $M_{\Sigma K}$, which prevents the existence of such a $p\phi$ pentaquark state with $J^P=\frac{1}{2}^-$. Our results also do not support the existence of bound states for $p\eta^\prime$, $\Lambda K$, $\Sigma K$ and $\Sigma^{\ast}K^{\ast}$ channels with $J^P=\frac{1}{2}^-$ and $\Sigma K^{\ast}$, $\Sigma^{\ast}K$ channels with $J^P=\frac{3}{2}^-$, because the obtained hadron masses are much higher than the corresponding two-hadron thresholds.
\begin{table}[]
  \caption{Extracted hadron masses of the hidden-strange molecular pentaquark states with $J^P=\frac{1}{2}^-$ and $J^P = \frac{3}{2}^-$.}
  \renewcommand\arraystretch{1.8} 
  \begin{tabular}{cccccc}
  \hline\hline
    Current  & Structure      & $J^P$    & $m_X$/GeV    &Threshold $s_0$/GeV$^2$    & Borel window/GeV$^2$      \\
    \hline
    $\eta_1(x)$ & $[p\eta^\prime]$        &$\frac{1}{2}^-$    & $1.91\pm 0.06 \pm 0.19$    & 4.77                      & 2.35-2.48                 \\
      $\eta_2(x)$ & $[p\phi]$               &$\frac{1}{2}^-$    & $1.95\pm 0.10 \pm 0.11$    & 4.62                      & 2.15-2.35                 \\
      $\eta_3(x)$ & $[\Lambda K]$           &$\frac{1}{2}^-$    & $2.93\pm 0.16 \pm 0.20$    & 10.57                     & 2.10-3.51                 \\
      $\eta_4(x)$ & $[\Sigma K]$            &$\frac{1}{2}^-$    & $2.63\pm 0.14 \pm 0.13$    & 8.35                      & 2.32-2.95                 \\
      $\eta_5(x)$ & $[\Sigma^\ast K^\ast]$  &$\frac{1}{2}^-$    & $2.78\pm 0.08 \pm 0.12$    & 9.60                      & 2.47-3.32                 \\
      $\eta_{6\mu}(x)$ & $[\Sigma K^*]$     &$\frac{3}{2}^-$    & $2.39\pm 0.08 \pm 0.11$    & 6.95                      & 2.47-3.32                 \\
      $\eta_{7\mu}(x)$ & $[\Sigma^* K]$     &$\frac{3}{2}^-$    & $2.27\pm 0.10 \pm 0.10$    & 5.82                      & 3.06-3.62                 \\
    \hline\hline         
  \end{tabular} \label{tablemass}
\end{table}

\section{Summary}
Inspired by the observations of hidden-charm $P_c$ states, we have investigated the hidden-strange pentaquark states in both $[udu][\bar ss]$ and $[uds][\bar su]$ configurations with quantum numbers $J^P=\frac{1}{2}^-$ and $J^P = \frac{3}{2}^-$ via QCD sum rule method. We have constructed the corresponding interpolating currents by using the $L=0$ baryonic and mesonic operators. After calculating the two-point correlation functions up to dimension-11, we perform the mass sum rule analyses by using the invariant structures proportional to $\slashed q$, and extracted the mass spectra for the $p\eta^\prime$, $p\phi$, $\Lambda K$, $\Sigma K$ and $\Sigma^\ast K^\ast$ pentaquarks with $J^P=\frac{1}{2}^-$ and $\Sigma K^\ast$ and $\Sigma^\ast K$ with $J^P= \frac{3}{2}^-$. 

There are only the recombination of quarks before and after strong decay, in that all pentaquarks with $J^P = \frac{1}{2}^-$ can decay into $\Sigma$ and $K$. Our calculations show that the masses for all $p\eta^\prime$, $p\phi$, $\Lambda K$, $\Sigma K$ and $\Sigma^\ast K^\ast$ pentaquarks are much higher than their two-hadron decay thresholds, so that there is no bound hidden-strange pentaquark state in these channels. 

For states $\Sigma^\ast K$ and $\Sigma K^\ast$ with $J^P = \frac{3}{2}^-$, they can decay into corresponding components $\Sigma^{(\ast)}$$K^{(\ast)}$. Our calculations show that these states also higher than mass thresholds corresponding components. Therefore there is also no bound hidden-strange pentaquarks state in such channels.This result is also consistent with the experimental status for the hidden-strange $P_s$ states~\cite{Belle:2017tfw}. 

\section*{ACKNOWLEDGMENTS}
This work is supported by the National Natural Science Foundation of China under Grant No. 12175318 and the National Key R$\&$D Program of China under Contracts No. 2020YFA0406400, and the Fundamental Research Funds for the Central Universities.


\appendix
\section{Appendix}
    \label{ros}
    In this appendix, we show the spectral densities for interpolating currents $\eta_2(x)$, $\eta_3(x)$, $\eta_4(x)$ and $\eta_5(x)$, including both $\rho_i^{q}(s)$ and $\rho_i^{m}(s)$ proportional to $\slashed{q}$ and 1 structures, respectively.

\begin{align}
    \begin{autobreak}
        \rho_2^q=
        \frac{s^5}{17203200 \pi ^7}
        +\frac{s^3 \gg }{983040 \pi ^7}
        +\frac{s^2 \qq{u} ^2}{1536 \pi ^3}
        +\frac{s^2 \qq{s} ^2}{1536 \pi ^3}
        -\frac{s^2 m_s \qgq{s} }{12288 \pi ^5}
        +\frac{g_s^2 s^2 \qq{u} ^2}{41472 \pi ^5}
        +\frac{g_s^2 s^2 \qq{s} ^2}{41472 \pi ^5}
        +\frac{g_s^2 s^2 \qq{d} ^2}{82944 \pi ^5}
        -\frac{s m_s \qq{s}  \gg }{18432 \pi ^5}
        +\frac{s \qq{s}  \qgq{s} }{384 \pi ^3}
        +\frac{s \qq{u}  \qgq{u} }{768 \pi ^3}
        -\frac{m_s \gg  \qgq{s} }{24576 \pi ^5}
        +\frac{5 \qq{s} ^2 \gg }{9216 \pi ^3}
        +\frac{\qq{u} ^2 \gg }{4608 \pi ^3}
        +\frac{g_s^3 \qq{s} ^2 \gg }{82944 \pi ^5}
        +\frac{g_s^3 \qq{u} ^2 \gg }{165888 \pi ^5}
        +\frac{g_s^2 m_s \qq{s} ^3}{1728 \pi ^3}
        +\frac{\qgq{s} ^2}{1024 \pi ^3}\, ,
    \end{autobreak}
\end{align}

\begin{align}
    \begin{autobreak}
        \rho_3^q=
        \frac{11 s^5}{825753600 \pi ^7}   
        +\frac{3 s^3 \gg }{5242880 \pi ^7}
        -\frac{s^3 m_s \qq{u} }{73728 \pi ^5}
        +\frac{11 s^3 m_s \qq{s} }{737280 \pi ^5}
        -\frac{s^2 \qq{u} ^2}{36864 \pi ^3}
        +\frac{5 s^2 \qq{s}  \qq{u} }{18432 \pi ^3}
        -\frac{13 s^2 m_s \qgq{s} }{1179648 \pi ^5}
        -\frac{13 s^2 m_s \qgq{u} }{196608 \pi ^5}    
        +\frac{11 g_s^2 s^2 \qq{u} ^2}{1990656 \pi ^5}
        +\frac{11 g_s^2 s^2 \qq{s} ^2}{1990656 \pi ^5}  
        +\frac{11 g_s^2 s^2 \qq{d} ^2}{3981312 \pi ^5}  
        -\frac{5 s m_s \qq{u}  \gg }{110592 \pi ^5}
        +\frac{35 s m_s \qq{s}  \gg }{589824 \pi ^5}
        +\frac{13 s \qq{u}  \qgq{s} }{18432 \pi ^3}
        +\frac{13 s \qq{s}  \qgq{u} }{18432 \pi ^3}
        +\frac{s \qq{u}  \qgq{u} }{18432 \pi ^3}
        +\frac{g_s^3 \qq{d} ^2 \gg }{442368 \pi ^5}
        -\frac{5 g_s^2 m_s \qq{d} ^2 \qq{u} }{62208 \pi ^3}
        +\frac{11 g_s^2 m_s \qq{d} ^2 \qq{s} }{124416 \pi ^3}
        +\frac{35 m_s \gg  \qgq{s} }{1179648 \pi ^5}
        -\frac{5 m_s \gg  \qgq{u} }{196608 \pi ^5}
        +\frac{25 \qq{s}  \qq{u}  \gg }{110592 \pi ^3}
        -\frac{5 \qq{u} ^2 \gg }{221184 \pi ^3}
        +\frac{35 g_s^3 \qq{s} ^2 \gg }{7962624 \pi ^5}
        +\frac{7 g_s^3 \qq{u} ^2 \gg }{1990656 \pi ^5}
        +\frac{\qgq{s}  \qgq{u} }{1536 \pi ^3}
        +\frac{\qgq{u} ^2}{12288 \pi ^3}
        -\frac{5 g_s^2 m_s \qq{u} ^3}{62208 \pi ^3}
        +\frac{11 g_s^2 m_s \qq{s}  \qq{u} ^2}{62208 \pi ^3}
        -\frac{5 g_s^2 m_s \qq{s} ^2 \qq{u} }{124416 \pi ^3}
        +\frac{11 g_s^2 m_s \qq{s} ^3}{124416 \pi ^3}
        -\frac{m_s \qq{s}  \qq{u} ^2}{1152 \pi }
        +\frac{5 m_s \qq{s} ^2 \qq{u} }{1152 \pi }\, ,
    \end{autobreak}
\end{align}

\begin{align}
    \begin{autobreak}
        \rho_4^q=
        \frac{11 s^5}{412876800 \pi ^7}
        +\frac{7 \gg  s^3}{5898240 \pi ^7}
        +\frac{\qq{d}  m_s s^3}{737280 \pi ^5}
        +\frac{11 \qq{s}  m_s s^3}{368640 \pi ^5}
        -\frac{\qq{u}  m_s s^3}{36864 \pi ^5}
        +\frac{11 g_s^2 \qq{d} ^2 s^2}{1990656 \pi ^5}
        +\frac{11 g_s^2 \qq{s} ^2 s^2}{995328 \pi ^5}
        -\frac{\qq{u} ^2 s^2}{18432 \pi ^3}
        +\frac{11 g_s^2 \qq{u} ^2 s^2}{995328 \pi ^5}
        -\frac{\qq{d}  \qq{s}  s^2}{36864 \pi ^3}
        +\frac{5 \qq{d}  \qq{u}  s^2}{18432 \pi ^3}
        +\frac{5 \qq{s}  \qq{u}  s^2}{9216 \pi ^3}
        +\frac{\qgq{d}  m_s s^2}{131072 \pi ^5}
        -\frac{89 \qgq{s}  m_s s^2}{1179648 \pi ^5}
        -\frac{25 \qgq{u}  m_s s^2}{196608 \pi ^5}
        -\frac{\qgq{d}  \qq{s}  s}{12288 \pi ^3}
        +\frac{\qq{d}  \qgq{s}  s}{12288 \pi ^3}
        +\frac{\qgq{d}  \qq{u}  s}{6144 \pi ^3}
        +\frac{25 \qgq{s}  \qq{u}  s}{18432 \pi ^3}
        +\frac{\qq{d}  \qgq{u}  s}{2048 \pi ^3}
        +\frac{25 \qq{s}  \qgq{u}  s}{18432 \pi ^3}
        +\frac{\qq{u}  \qgq{u}  s}{18432 \pi ^3}
        +\frac{13 \gg  \qq{d}  m_s s}{884736 \pi ^5}
        +\frac{53 \gg  \qq{s}  m_s s}{589824 \pi ^5}
        -\frac{5 \gg  \qq{u}  m_s s}{55296 \pi ^5}
        +\frac{5 g_s^3 \gg  \qq{d} ^2}{884736 \pi ^5}
        +\frac{53 g_s^3 \gg  \qq{s} ^2}{7962624 \pi ^5}
        -\frac{5 \gg  \qq{u} ^2}{110592 \pi ^3}
        +\frac{35 g_s^3 \gg  \qq{u} ^2}{3981312 \pi ^5}
        +\frac{\qgq{u} ^2}{8192 \pi ^3}
        -\frac{7 \gg  \qq{d}  \qq{s} }{110592 \pi ^3}
        +\frac{\qgq{d}  \qgq{s} }{24576 \pi ^3}
        +\frac{17 \gg  \qq{d}  \qq{u} }{55296 \pi ^3}
        +\frac{25 \gg  \qq{s}  \qq{u} }{55296 \pi ^3}
        +\frac{\qgq{d}  \qgq{u} }{12288 \pi ^3}
        +\frac{5 \qgq{s}  \qgq{u} }{4096 \pi ^3}
        +\frac{11 g_s^2 \qq{s} ^3 m_s}{62208 \pi ^3}
        -\frac{5 g_s^2 \qq{u} ^3 m_s}{31104 \pi ^3}
        -\frac{\qq{d}  \qq{s} ^2 m_s}{2304 \pi }
        +\frac{g_s^2 \qq{d}  \qq{s} ^2 m_s}{248832 \pi ^3}
        -\frac{11 \qq{d}  \qq{u} ^2 m_s}{1152 \pi }
        +\frac{g_s^2 \qq{d}  \qq{u} ^2 m_s}{62208 \pi ^3}
        -\frac{\qq{s}  \qq{u} ^2 m_s}{576 \pi }
        +\frac{11 g_s^2 \qq{s}  \qq{u} ^2 m_s}{31104 \pi ^3}
        +\frac{\gg  \qgq{d}  m_s}{98304 \pi ^5}
        +\frac{11 g_s^2 \qq{d} ^2 \qq{s}  m_s}{62208 \pi ^3}
        +\frac{53 \gg  \qgq{s}  m_s}{1179648 \pi ^5}
        -\frac{5 g_s^2 \qq{d} ^2 \qq{u}  m_s}{31104 \pi ^3
        }+\frac{5 \qq{s} ^2 \qq{u}  m_s}{576 \pi }
        -\frac{5 g_s^2 \qq{s} ^2 \qq{u}  m_s}{62208 \pi ^3}
        +\frac{5 \qq{d}  \qq{s}  \qq{u}  m_s}{576 \pi }
        -\frac{5 \gg  \qgq{u}  m_s}{98304 \pi ^5}\, ,       
    \end{autobreak}
\end{align}

\begin{align}
    \begin{autobreak}
        \rho_5^q = 
        \frac{3 s^5}{11468800 \pi ^7}
        -\frac{\gg  s^3}{655360 \pi ^7}
        -\frac{3 \qq{d}  m_s s^3}{20480 \pi ^5}
        +\frac{3 \qq{s}  m_s s^3}{10240 \pi ^5}
        -\frac{\qq{u}  m_s s^3}{5120 \pi ^5}
        +\frac{g_s^2 \qq{d} ^2 s^2}{18432 \pi ^5}
        +\frac{g_s^2 \qq{s} ^2 s^2}{9216 \pi ^5}
        +\frac{\qq{u} ^2 s^2}{512 \pi ^3}
        +\frac{g_s^2 \qq{u} ^2 s^2}{9216 \pi ^5}
        +\frac{3 \qq{d}  \qq{s}  s^2}{1024 \pi ^3}
        +\frac{\qq{s}  \qq{u}  s^2}{256 \pi ^3}
        -\frac{9 \qgq{d}  m_s s^2}{16384 \pi ^5}
        +\frac{7 \qgq{s}  m_s s^2}{8192 \pi ^5}
        -\frac{7 \qgq{u}  m_s s^2}{8192 \pi ^5}
        +\frac{3 \qgq{d}  \qq{s}  s}{512 \pi ^3}
        +\frac{3 \qq{d}  \qgq{s}  s}{512 \pi ^3}
        +\frac{7 \qgq{s}  \qq{u}  s}{768 \pi ^3}
        +\frac{7 \qq{s}  \qgq{u}  s}{768 \pi ^3}
        +\frac{7 \qq{u}  \qgq{u}  s}{768 \pi ^3}
        +\frac{\gg  \qq{d}  m_s s}{2048 \pi ^5}
        -\frac{\gg  \qq{s}  m_s s}{6144 \pi ^5}
        +\frac{\gg  \qq{u}  m_s s}{2048 \pi ^5}
        -\frac{g_s^3 \gg  \qq{d} ^2}{110592 \pi ^5}
        -\frac{g_s^3 \gg  \qq{s} ^2}{82944 \pi ^5}
        -\frac{\gg  \qq{u} ^2}{1536 \pi ^3}
        -\frac{g_s^3 \gg  \qq{u} ^2}{165888 \pi ^5}
        +\frac{\qgq{u} ^2}{256 \pi ^3}
        -\frac{3 \gg  \qq{d}  \qq{s} }{2048 \pi ^3}
        +\frac{9 \qgq{d}  \qgq{s} }{2048 \pi ^3}
        -\frac{\gg  \qq{s}  \qq{u} }{768 \pi ^3}
        +\frac{\qgq{s}  \qgq{u} }{128 \pi ^3}
        +\frac{g_s^2 \qq{s} ^3 m_s}{576 \pi ^3}
        -\frac{g_s^2 \qq{u} ^3 m_s}{864 \pi ^3}
        +\frac{3 \qq{d}  \qq{s} ^2 m_s}{64 \pi }
        -\frac{g_s^2 \qq{d}  \qq{s} ^2 m_s}{2304 \pi ^3}
        -\frac{3 \qq{d}  \qq{u} ^2 m_s}{32 \pi }
        -\frac{g_s^2 \qq{d}  \qq{u} ^2 m_s}{576 \pi ^3}
        +\frac{\qq{s}  \qq{u} ^2 m_s}{16 \pi }
        +\frac{g_s^2 \qq{s}  \qq{u} ^2 m_s}{288 \pi ^3}
        +\frac{15 \gg  \qgq{d}  m_s}{32768 \pi ^5}
        +\frac{g_s^2 \qq{d} ^2 \qq{s}  m_s}{576 \pi ^3}
        -\frac{\gg  \qgq{s}  m_s}{12288 \pi ^5}
        -\frac{g_s^2 \qq{d} ^2 \qq{u}  m_s}{864 \pi ^3}
        +\frac{\qq{s} ^2 \qq{u}  m_s}{16 \pi }
        -\frac{g_s^2 \qq{s} ^2 \qq{u}  m_s}{1728 \pi ^3}
        -\frac{3 \qq{d}  \qq{s}  \qq{u}  m_s}{8 \pi }
        +\frac{\gg  \qgq{u}  m_s}{2048 \pi ^5}\, ,
    \end{autobreak}
\end{align}

\begin{align}
    \begin{autobreak}
        \rho_{6}^{q}=
        \frac{31 s^5}{440401920 \pi ^7}
        -\frac{11 \left\langle g_s^2 G^2 \right\rangle  s^3}{283115520 \pi ^7}
        +\frac{\qq{d}  m_s s^3}{245760 \pi ^5}
        +\frac{37 \qq{s}  m_s s^3}{491520 \pi ^5}
        -\frac{\qq{u}  m_s s^3}{12288 \pi ^5}
        +\frac{89 g_s^2 \qq{d} ^2 s^2}{6635520 \pi ^5}
        +\frac{89 g_s^2 \qq{s} ^2 s^2}{3317760 \pi ^5}
        -\frac{3 \qq{u} ^2 s^2}{20480 \pi ^3}
        +\frac{89 g_s^2 \qq{u} ^2 s^2}{3317760 \pi ^5}
        -\frac{\qq{d}  \qq{s}  s^2}{12288 \pi ^3}
        +\frac{\qq{d}  \qq{u}  s^2}{1536 \pi ^3}
        +\frac{5 \qq{s}  \qq{u}  s^2}{3072 \pi ^3}
        +\frac{3 \qgq{d}  m_s s^2}{131072 \pi ^5}
        +\frac{65 \qgq{s}  m_s s^2}{393216 \pi ^5}
        -\frac{25 \qgq{u}  m_s s^2}{65536 \pi ^5}
        -\frac{\qgq{d}  \qq{s}  s}{4096 \pi ^3}
        -\frac{\qq{d}  \qgq{s}  s}{4608 \pi ^3}
        +\frac{13 \qgq{d}  \qq{u}  s}{24576 \pi ^3}
        +\frac{113 \qgq{s}  \qq{u}  s}{36864 \pi ^3}
        +\frac{17 \qq{d}  \qgq{u}  s}{24576 \pi ^3}
        +\frac{25 \qq{s}  \qgq{u}  s}{6144 \pi ^3}
        -\frac{25 \qq{u}  \qgq{u}  s}{36864 \pi ^3}
        +\frac{13 \left\langle g_s^2 G^2 \right\rangle  \qq{d}  m_s s}{294912 \pi ^5}
        -\frac{19 \left\langle g_s^2 G^2 \right\rangle  \qq{s}  m_s s}{589824 \pi ^5}
        -\frac{5 \left\langle g_s^2 G^2 \right\rangle  \qq{u}  m_s s}{18432 \pi ^5}
        -\frac{59 g_s^3 \left\langle g_s^2 G^2 \right\rangle  \qq{d} ^2}{11943936 \pi ^5}
        -\frac{25 g_s^3 \left\langle g_s^2 G^2 \right\rangle  \qq{s} ^2}{7962624 \pi ^5}
        -\frac{5 \left\langle g_s^2 G^2 \right\rangle  \qq{u} ^2}{73728 \pi ^3}
        +\frac{13 g_s^3 \left\langle g_s^2 G^2 \right\rangle  \qq{u} ^2}{5971968 \pi ^5}
        -\frac{11 \qgq{u} ^2}{36864 \pi ^3}
        -\frac{7 \left\langle g_s^2 G^2 \right\rangle  \qq{d}  \qq{s} }{36864 \pi ^3}
        -\frac{5 \qgq{d}  \qgq{s} }{24576 \pi ^3}
        -\frac{\left\langle g_s^2 G^2 \right\rangle  \qq{d}  \qq{u} }{4608 \pi ^3}
        +\frac{25 \left\langle g_s^2 G^2 \right\rangle  \qq{s}  \qq{u} }{18432 \pi ^3}
        -\frac{\qgq{d}  \qgq{u} }{36864 \pi ^3}
        +\frac{3 \qgq{s}  \qgq{u} }{1024 \pi ^3}
        +\frac{5 g_s^2 \qq{s} ^3 m_s}{13824 \pi ^3}
        -\frac{5 g_s^2 \qq{u} ^3 m_s}{10368 \pi ^3}
        -\frac{\qq{d}  \qq{s} ^2 m_s}{768 \pi }
        +\frac{g_s^2 \qq{d}  \qq{s} ^2 m_s}{82944 \pi ^3}
        -\frac{11 \qq{d}  \qq{u} ^2 m_s}{384 \pi }
        +\frac{g_s^2 \qq{d}  \qq{u} ^2 m_s}{20736 \pi ^3}
        -\frac{5 \qq{s}  \qq{u} ^2 m_s}{1152 \pi }
        +\frac{5 g_s^2 \qq{s}  \qq{u} ^2 m_s}{6912 \pi ^3}
        +\frac{\left\langle g_s^2 G^2 \right\rangle  \qgq{d}  m_s}{32768 \pi ^5}
        +\frac{5 g_s^2 \qq{d} ^2 \qq{s}  m_s}{13824 \pi ^3}
        -\frac{25 \left\langle g_s^2 G^2 \right\rangle  \qgq{s}  m_s}{1179648 \pi ^5}
        -\frac{5 g_s^2 \qq{d} ^2 \qq{u}  m_s}{10368 \pi ^3}
        +\frac{5 \qq{s} ^2 \qq{u}  m_s}{192 \pi }
        -\frac{5 g_s^2 \qq{s} ^2 \qq{u}  m_s}{20736 \pi ^3}
        +\frac{5 \qq{d}  \qq{s}  \qq{u}  m_s}{288 \pi }
        -\frac{5 \left\langle g_s^2 G^2 \right\rangle  \qgq{u}  m_s}{32768 \pi ^5}
        +\, ,
    \end{autobreak}
\end{align}

\begin{align}
    \begin{autobreak}
        \rho^q_{7} = 
        \frac{3 s^5}{18350080 \pi ^7}
        +\frac{19 \left\langle g_s^2 G^2 \right\rangle  s^3}{7864320 \pi ^7}
        -\frac{7 \qq{d}  m_s s^3}{40960 \pi ^5}
        +\frac{7 \qq{s}  m_s s^3}{40960 \pi ^5}
        -\frac{\qq{u}  m_s s^3}{4096 \pi ^5}
        +\frac{11 g_s^2 \qq{d} ^2 s^2}{368640 \pi ^5}
        +\frac{11 g_s^2 \qq{s} ^2 s^2}{184320 \pi ^5}
        +\frac{5 \qq{u} ^2 s^2}{2048 \pi ^3}
        +\frac{11 g_s^2 \qq{u} ^2 s^2}{184320 \pi ^5}
        +\frac{33 \qq{d}  \qq{s}  s^2}{10240 \pi ^3}+\frac{5 \qq{s}  \qq{u}  s^2}{1024 \pi ^3}
        -\frac{99 \qgq{d}  m_s s^2}{163840 \pi ^5}
        +\frac{3 \qgq{s}  m_s s^2}{81920 \pi ^5}
        -\frac{35 \qgq{u}  m_s s^2}{32768 \pi ^5}
        +\frac{3 \qgq{d}  \qq{s}  s}{512 \pi ^3}
        +\frac{3 \qq{d}  \qgq{s}  s}{512 \pi ^3}
        +\frac{35 \qgq{s}  \qq{u}  s}{3072 \pi ^3}
        +\frac{35 \qq{s}  \qgq{u}  s}{3072 \pi ^3}
        +\frac{35 \qq{u}  \qgq{u}  s}{3072 \pi ^3}
        +\frac{3 \left\langle g_s^2 G^2 \right\rangle  \qq{d}  m_s s}{8192 \pi ^5}
        +\frac{\left\langle g_s^2 G^2 \right\rangle  \qq{s}  m_s s}{24576 \pi ^5}
        -\frac{35 \left\langle g_s^2 G^2 \right\rangle  \qq{u}  m_s s}{24576 \pi ^5}
        -\frac{25 g_s^3 \left\langle g_s^2 G^2 \right\rangle  \qq{d} ^2}{1327104 \pi ^5}
        -\frac{5 g_s^3 \left\langle g_s^2 G^2 \right\rangle  \qq{s} ^2}{1990656 \pi ^5}
        +\frac{5 \left\langle g_s^2 G^2 \right\rangle  \qq{u} ^2}{1536 \pi ^3}
        +\frac{65 g_s^3 \left\langle g_s^2 G^2 \right\rangle  \qq{u} ^2}{1990656 \pi ^5}
        +\frac{5 \qgq{u} ^2}{1024 \pi ^3}
        -\frac{5 \left\langle g_s^2 G^2 \right\rangle  \qq{d}  \qq{s} }{4096 \pi ^3}
        +\frac{15 \qgq{d}  \qgq{s} }{4096 \pi ^3}
        +\frac{5 \left\langle g_s^2 G^2 \right\rangle  \qq{s}  \qq{u} }{768 \pi ^3}
        +\frac{5 \qgq{s}  \qgq{u} }{512 \pi ^3}
        +\frac{5 g_s^2 \qq{s} ^3 m_s}{6912 \pi ^3}
        -\frac{5 g_s^2 \qq{u} ^3 m_s}{3456 \pi ^3}
        +\frac{5 \qq{d}  \qq{s} ^2 m_s}{128 \pi }
        -\frac{5 g_s^2 \qq{d}  \qq{s} ^2 m_s}{13824 \pi ^3}
        -\frac{5 \qq{d}  \qq{u} ^2 m_s}{64 \pi }
        -\frac{5 g_s^2 \qq{d}  \qq{u} ^2 m_s}{3456 \pi ^3}
        +\frac{5 \qq{s}  \qq{u} ^2 m_s}{64 \pi }
        +\frac{5 g_s^2 \qq{s}  \qq{u} ^2 m_s}{3456 \pi ^3}
        +\frac{25 \left\langle g_s^2 G^2 \right\rangle  \qgq{d}  m_s}{65536 \pi ^5}
        +\frac{5 g_s^2 \qq{d} ^2 \qq{s}  m_s}{6912 \pi ^3}
        -\frac{5 \left\langle g_s^2 G^2 \right\rangle  \qgq{s}  m_s}{294912 \pi ^5}
        -\frac{5 g_s^2 \qq{d} ^2 \qq{u}  m_s}{3456 \pi ^3}
        +\frac{5 \qq{s} ^2 \qq{u}  m_s}{64 \pi }
        -\frac{5 g_s^2 \qq{s} ^2 \qq{u}  m_s}{6912 \pi ^3}
        -\frac{5 \qq{d}  \qq{s}  \qq{u}  m_s}{16 \pi }
        -\frac{15 \left\langle g_s^2 G^2 \right\rangle  \qgq{u}  m_s}{16384 \pi ^5}
        +\, .
    \end{autobreak}
\end{align}

The spectral densities $\rho_i^{m}$ are

\begin{align}
    \begin{autobreak}
        \rho_2^m =
        \frac{s^4 \qq{d} }{245760 \pi ^5}
        -\frac{s^2 m_s \qq{d}  \qq{s} }{256 \pi ^3}
        -\frac{s^2 \qq{d}  \gg }{18432 \pi ^5}
        +\frac{s \qq{d}  \qq{u} ^2}{48 \pi }
        +\frac{s \qq{d}  \qq{s} ^2}{48 \pi }
        -\frac{3 s \gg  \qgq{d} }{32768 \pi ^5}
        -\frac{5 s m_s \qq{d}  \qgq{s} }{768 \pi ^3}
        +\frac{g_s^2 s \qq{d}  \qq{u} ^2}{5184 \pi ^3}
        +\frac{g_s^2 s \qq{d}  \qq{s} ^2}{5184 \pi ^3}
        +\frac{m_s \qq{d}  \qq{s}  \gg }{4608 \pi ^3}
        -\frac{\qq{d}  \gg ^2}{196608 \pi ^5}
        +\frac{g_s^2 \qq{u} ^2 \qgq{d} }{20736 \pi ^3}
        +\frac{m_s \qgq{d}  \qgq{s} }{512 \pi ^3}
        +\frac{\qq{d}  \qq{s}  \qgq{s} }{48 \pi }
        +\frac{\qq{u} ^2 \qgq{d} }{96 \pi }
        +\frac{\qq{d}  \qq{u}  \qgq{u} }{48 \pi }\, ,
    \end{autobreak}
\end{align}

\begin{align}
    \begin{autobreak}
        \rho_3^m=
        \frac{5 s^2 m_s \qq{d}  \qq{u} }{4608 \pi ^3}
        -\frac{s^2 m_s \qq{d}  \qq{s} }{4608 \pi ^3}
        -\frac{5 s \qq{d}  \qq{s}  \qq{u} }{576 \pi }
        +\frac{s \qq{d}  \qq{s} ^2}{1152 \pi }
        \frac{5 m_s \qq{d}  \qq{u}  \gg }{6912 \pi ^3}
        -\frac{m_s \qq{d}  \qq{s}  \gg }{6912 \pi ^3}
        +\frac{m_s \qgq{d}  \qgq{s} }{6144 \pi ^3}
        +\frac{s m_s \qq{u}  \qgq{d} }{512 \pi ^3}
        +\frac{7 s m_s \qq{d}  \qgq{u} }{3072 \pi ^3}
        +\frac{m_s \qgq{d}  \qgq{u} }{768 \pi ^3}
        -\frac{\qq{s}  \qq{u}  \qgq{d} }{192 \pi }
        -\frac{5 \qq{d}  \qq{u}  \qgq{s} }{1152 \pi }
        -\frac{7 \qq{d}  \qq{s}  \qgq{u} }{1152 \pi }\, ,
    \end{autobreak}
\end{align}

\begin{align}
    \begin{autobreak}
        \rho_4^m=
        \frac{11 m_s s^5}{117964800 \pi ^7}
        -\frac{11 \qq{s}  s^4}{2949120 \pi ^5}
        -\frac{\qgq{s}  s^3}{589824 \pi ^5}
        +\frac{\gg  m_s s^3}{1048576 \pi ^7}
        -\frac{19 \gg  \qq{s}  s^2}{884736 \pi ^5}
        +\frac{11 g_s^2 \qq{d} ^2 m_s s^2}{995328 \pi ^5}-\frac{11 \qq{s} ^2 m_s s^2}{18432 \pi ^3}
        +\frac{11 g_s^2 \qq{s} ^2 m_s s^2}{1990656 \pi ^5}-\frac{\qq{u} ^2 m_s s^2}{9216 \pi ^3}
        +\frac{11 g_s^2 \qq{u} ^2 m_s s^2}{497664 \pi ^5}
        -\frac{\qq{d}  \qq{s}  m_s s^2}{2304 \pi ^3}
        +\frac{5 \qq{d}  \qq{u}  m_s s^2}{2304 \pi ^3}
        +\frac{5 \qq{s}  \qq{u}  m_s s^2}{2304 \pi ^3}
        -\frac{11 g_s^2 \qq{s} ^3 s}{124416 \pi ^3}
        +\frac{\qq{d}  \qq{s} ^2 s}{576 \pi }
        +\frac{\qq{s}  \qq{u} ^2 s}{1152 \pi }
        -\frac{11 g_s^2 \qq{s}  \qq{u} ^2 s}{62208 \pi ^3}
        -\frac{11 g_s^2 \qq{d} ^2 \qq{s}  s}{124416 \pi ^3}
        -\frac{3 \gg  \qgq{s}  s}{131072 \pi ^5}
        -\frac{5 \qq{s} ^2 \qq{u}  s}{576 \pi }
        -\frac{5 \qq{d}  \qq{s}  \qq{u}  s}{288 \pi }
        -\frac{\qgq{d}  \qq{s}  m_s s}{1536 \pi ^3}
        +\frac{\qq{d}  \qgq{s}  m_s s}{6144 \pi ^3}
        -\frac{47 \qq{s}  \qgq{s}  m_s s}{73728 \pi ^3}
        +\frac{5 \qgq{d}  \qq{u}  m_s s}{1536 \pi ^3}
        +\frac{11 \qgq{s}  \qq{u}  m_s s}{6144 \pi ^3}
        +\frac{7 \qq{d}  \qgq{u}  m_s s}{1536 \pi ^3}
        +\frac{11 \qq{s}  \qgq{u}  m_s s}{3072 \pi ^3}
        -\frac{\qq{u}  \qgq{u}  m_s s}{6144 \pi ^3}
        +\frac{\qgq{d}  \qq{s} ^2}{1152 \pi }
        -\frac{7 g_s^2 \qgq{s}  \qq{u} ^2}{248832 \pi ^3}
        -\frac{\gg ^2 \qq{s} }{786432 \pi ^5}
        -\frac{13 g_s^2 \qq{d} ^2 \qgq{s} }{497664 \pi ^3}
        -\frac{g_s^2 \qq{s} ^2 \qgq{s} }{497664 \pi ^3}
        -\frac{\qq{d}  \qq{s}  \qgq{s} }{2304 \pi }
        -\frac{5 \qgq{d}  \qq{s}  \qq{u} }{576 \pi }
        -\frac{11 \qq{d}  \qgq{s}  \qq{u} }{1152 \pi }
        -\frac{11 \qq{s}  \qgq{s}  \qq{u} }{2304 \pi }
        -\frac{11 \qq{s} ^2 \qgq{u} }{2304 \pi }
        -\frac{7 \qq{d}  \qq{s}  \qgq{u} }{576 \pi }
        +\frac{\qq{s}  \qq{u}  \qgq{u} }{2304 \pi }
        +\frac{g_s^3 \gg  \qq{d} ^2 m_s}{221184 \pi ^5}
        +\frac{25 \gg  \qq{s} ^2 m_s}{221184 \pi ^3}
        -\frac{11 g_s^3 \gg  \qq{s} ^2 m_s}{2654208 \pi ^5}
        +\frac{31 \qgq{s} ^2 m_s}{36864 \pi ^3}
        +\frac{\gg  \qq{u} ^2 m_s}{110592 \pi ^3}
        +\frac{g_s^3 \gg  \qq{u} ^2 m_s}{331776 \pi ^5}
        -\frac{\gg  \qq{d}  \qq{s}  m_s}{3456 \pi ^3}
        +\frac{\qgq{d}  \qgq{s}  m_s}{12288 \pi ^3}
        +\frac{5 \gg  \qq{d}  \qq{u}  m_s}{3456 \pi ^3}
        -\frac{5 \gg  \qq{s}  \qq{u}  m_s}{27648 \pi ^3}
        +\frac{7 \qgq{d}  \qgq{u}  m_s}{3072 \pi ^3}
        +\frac{13 \qgq{s}  \qgq{u}  m_s}{12288 \pi ^3}\, ,
    \end{autobreak}
\end{align}

\begin{align}
    \begin{autobreak}
        \rho_5^m=
        \frac{3 m_s s^5}{3276800 \pi ^7}
        -\frac{3 \qq{s}  s^4}{81920 \pi ^5}
        -\frac{3 \qq{u}  s^4}{40960 \pi ^5}
        -\frac{\qgq{s}  s^3}{4096 \pi ^5}
        -\frac{\qgq{u}  s^3}{2048 \pi ^5}
        -\frac{\gg  m_s s^3}{262144 \pi ^7}
        +\frac{g_s^2 \qq{d} ^2 m_s s^2}{9216 \pi ^5}
        -\frac{3 \qq{s} ^2 m_s s^2}{512 \pi ^3}
        +\frac{g_s^2 \qq{s} ^2 m_s s^2}{18432 \pi ^5}
        +\frac{3 \qq{u} ^2 m_s s^2}{256 \pi ^3}
        +\frac{g_s^2 \qq{u} ^2 m_s s^2}{4608 \pi ^5}
        +\frac{\qq{d}  \qq{s}  m_s s^2}{64 \pi ^3}
        +\frac{\qq{d}  \qq{u}  m_s s^2}{64 \pi ^3}
        -\frac{3 \qq{s}  \qq{u}  m_s s^2}{128 \pi ^3}
        -\frac{g_s^2 \qq{s} ^3 s}{1152 \pi ^3}
        -\frac{g_s^2 \qq{u} ^3 s}{576 \pi ^3}
        -\frac{\qq{d}  \qq{s} ^2 s}{16 \pi }
        -\frac{3 \qq{s}  \qq{u} ^2 s}{32 \pi }
        -\frac{g_s^2 \qq{s}  \qq{u} ^2 s}{576 \pi ^3}
        -\frac{g_s^2 \qq{d} ^2 \qq{s}  s}{1152 \pi ^3}
        +\frac{3 \gg  \qgq{s}  s}{32768 \pi ^5}
        -\frac{g_s^2 \qq{d} ^2 \qq{u}  s}{576 \pi ^3}
        -\frac{g_s^2 \qq{s} ^2 \qq{u}  s}{288 \pi ^3}
        -\frac{\qq{d}  \qq{s}  \qq{u}  s}{8 \pi }
        +\frac{3 \gg  \qgq{u}  s}{16384 \pi ^5}
        +\frac{3 \qgq{d}  \qq{s}  m_s s}{128 \pi ^3}
        +\frac{7 \qq{d}  \qgq{s}  m_s s}{256 \pi ^3}
        -\frac{9 \qq{s}  \qgq{s}  m_s s}{512 \pi ^3}
        +\frac{3 \qgq{d}  \qq{u}  m_s s}{128 \pi ^3}
        -\frac{3 \qgq{s}  \qq{u}  m_s s}{128 \pi ^3}
        +\frac{\qq{d}  \qgq{u}  m_s s}{32 \pi ^3}
        -\frac{45 \qq{s}  \qgq{u}  m_s s}{1024 \pi ^3}
        +\frac{9 \qq{u}  \qgq{u}  m_s s}{256 \pi ^3}
        -\frac{\qgq{d}  \qq{s} ^2}{32 \pi }
        -\frac{3 \qgq{s}  \qq{u} ^2}{64 \pi }
        -\frac{7 g_s^2 \qgq{s}  \qq{u} ^2}{6912 \pi ^3}
        +\frac{\gg ^2 \qq{s} }{196608 \pi ^5}
        -\frac{g_s^2 \qq{d} ^2 \qgq{s} }{1728 \pi ^3}
        -\frac{g_s^2 \qq{s} ^2 \qgq{s} }{1728 \pi ^3}
        -\frac{7 \qq{d}  \qq{s}  \qgq{s} }{96 \pi }
        +\frac{\gg ^2 \qq{u} }{98304 \pi ^5}
        -\frac{\qgq{d}  \qq{s}  \qq{u} }{16 \pi }
        -\frac{\qq{d}  \qgq{s}  \qq{u} }{16 \pi }
        -\frac{g_s^2 \qq{d} ^2 \qgq{u} }{864 \pi ^3}
        -\frac{5 g_s^2 \qq{s} ^2 \qgq{u} }{2304 \pi ^3}
        -\frac{7 g_s^2 \qq{u} ^2 \qgq{u} }{6912 \pi ^3}
        -\frac{\qq{d}  \qq{s}  \qgq{u} }{12 \pi }
        -\frac{3 \qq{s}  \qq{u}  \qgq{u} }{32 \pi }
        -\frac{g_s^3 \gg  \qq{d} ^2 m_s}{331776 \pi ^5}
        -\frac{\gg  \qq{s} ^2 m_s}{3072 \pi ^3}
        +\frac{g_s^3 \gg  \qq{s} ^2 m_s}{663552 \pi ^5}
        -\frac{5 \qgq{s} ^2 m_s}{1024 \pi ^3}
        -\frac{\gg  \qq{u} ^2 m_s}{256 \pi ^3}
        -\frac{g_s^3 \gg  \qq{u} ^2 m_s}{82944 \pi ^5}
        +\frac{9 \qgq{u} ^2 m_s}{1024 \pi ^3}
        -\frac{\gg  \qq{d}  \qq{s}  m_s}{768 \pi ^3}
        +\frac{7 \qgq{d}  \qgq{s}  m_s}{512 \pi ^3}
        -\frac{\gg  \qq{d}  \qq{u}  m_s}{768 \pi ^3}
        -\frac{\gg  \qq{s}  \qq{u}  m_s}{1024 \pi ^3}
        +\frac{\qgq{d}  \qgq{u}  m_s}{64 \pi ^3}
        -\frac{9 \qgq{s}  \qgq{u}  m_s}{512 \pi ^3}\, .
    \end{autobreak}
\end{align}

\begin{align}
    \begin{autobreak}
        \rho_{6}^m =
        \frac{11 m_s s^5}{45875200 \pi ^7}
        -\frac{11 \qq{s}  s^4}{1179648 \pi ^5}
        -\frac{\qgq{s}  s^3}{983040 \pi ^5}
        -\frac{3 \left\langle g_s^2 G^2 \right\rangle  m_s s^3}{524288 \pi ^7}
        +\frac{5 \left\langle g_s^2 G^2 \right\rangle  \qq{s}  s^2}{65536 \pi ^5}
        +\frac{11 g_s^2 \qq{d} ^2 m_s s^2}{442368 \pi ^5}
        -\frac{11 \qq{s} ^2 m_s s^2}{8192 \pi ^3}
        +\frac{11 g_s^2 \qq{s} ^2 m_s s^2}{884736 \pi ^5}
        -\frac{\qq{u} ^2 m_s s^2}{4096 \pi ^3}
        +\frac{11 g_s^2 \qq{u} ^2 m_s s^2}{221184 \pi ^5}
        -\frac{\qq{d}  \qq{s}  m_s s^2}{768 \pi ^3}
        +\frac{5 \qq{d}  \qq{u}  m_s s^2}{1024 \pi ^3}
        +\frac{5 \qq{s}  \qq{u}  m_s s^2}{768 \pi ^3}
        -\frac{11 g_s^2 \qq{s} ^3 s}{62208 \pi ^3}
        +\frac{\qq{d}  \qq{s} ^2 s}{192 \pi }
        +\frac{\qq{s}  \qq{u} ^2 s}{576 \pi }
        -\frac{11 g_s^2 \qq{s}  \qq{u} ^2 s}{31104 \pi ^3}
        -\frac{11 g_s^2 \qq{d} ^2 \qq{s}  s}{62208 \pi ^3}
        +\frac{13 \left\langle g_s^2 G^2 \right\rangle  \qgq{s}  s}{98304 \pi ^5}
        -\frac{5 \qq{s} ^2 \qq{u}  s}{192 \pi }
        -\frac{5 \qq{d}  \qq{s}  \qq{u}  s}{144 \pi }
        -\frac{\qgq{d}  \qq{s}  m_s s}{512 \pi ^3}
        -\frac{31 \qq{d}  \qgq{s}  m_s s}{18432 \pi ^3}
        -\frac{47 \qq{s}  \qgq{s}  m_s s}{36864 \pi ^3}
        +\frac{5 \qgq{d}  \qq{u}  m_s s}{768 \pi ^3}
        +\frac{89 \qgq{s}  \qq{u}  m_s s}{18432 \pi ^3}
        +\frac{\qq{d}  \qgq{u}  m_s s}{144 \pi ^3}
        +\frac{11 \qq{s}  \qgq{u}  m_s s}{1024 \pi ^3}
        -\frac{13 \qq{u}  \qgq{u}  m_s s}{9216 \pi ^3}
        +\frac{\qgq{d}  \qq{s} ^2}{384 \pi }
        +\frac{\qgq{s}  \qq{u} ^2}{1536 \pi }
        -\frac{5 g_s^2 \qgq{s}  \qq{u} ^2}{165888 \pi ^3}
        +\frac{11 \left\langle g_s^2 G^2 \right\rangle ^2 \qq{s} }{1572864 \pi ^5}
        -\frac{11 g_s^2 \qq{d} ^2 \qgq{s} }{331776 \pi ^3}
        -\frac{g_s^2 \qq{s} ^2 \qgq{s} }{331776 \pi ^3}
        +\frac{\qq{d}  \qq{s}  \qgq{s} }{256 \pi }
        -\frac{5 \qgq{d}  \qq{s}  \qq{u} }{384 \pi }
        -\frac{5 \qq{d}  \qgq{s}  \qq{u} }{384 \pi }
        -\frac{5 \qq{s}  \qgq{s}  \qq{u} }{384 \pi }
        -\frac{11 \qq{s} ^2 \qgq{u} }{768 \pi }
        -\frac{5 \qq{d}  \qq{s}  \qgq{u} }{384 \pi }
        +\frac{5 \qq{s}  \qq{u}  \qgq{u} }{1536 \pi }
        -\frac{11 g_s^3 \left\langle g_s^2 G^2 \right\rangle  \qq{d} ^2 m_s}{1327104 \pi ^5}
        +\frac{25 \left\langle g_s^2 G^2 \right\rangle  \qq{s} ^2 m_s}{147456 \pi ^3}
        +\frac{g_s^3 \left\langle g_s^2 G^2 \right\rangle  \qq{s} ^2 m_s}{589824 \pi ^5}
        -\frac{\qgq{s} ^2 m_s}{768 \pi ^3}
        -\frac{5 \left\langle g_s^2 G^2 \right\rangle  \qq{u} ^2 m_s}{73728 \pi ^3}
        -\frac{g_s^3 \left\langle g_s^2 G^2 \right\rangle  \qq{u} ^2 m_s}{82944 \pi ^5}
        -\frac{\qgq{u} ^2 m_s}{2048 \pi ^3}
        -\frac{\left\langle g_s^2 G^2 \right\rangle  \qq{d}  \qq{s}  m_s}{1152 \pi ^3}
        -\frac{3 \qgq{d}  \qgq{s}  m_s}{4096 \pi ^3}
        -\frac{5 \left\langle g_s^2 G^2 \right\rangle  \qq{d}  \qq{u}  m_s}{4608 \pi ^3}
        -\frac{5 \left\langle g_s^2 G^2 \right\rangle  \qq{s}  \qq{u}  m_s}{9216 \pi ^3}
        +\frac{5 \qgq{d}  \qgq{u}  m_s}{2048 \pi ^3}
        +\frac{3 \qgq{s}  \qgq{u}  m_s}{1024 \pi ^3}
        +\, ,
    \end{autobreak}
\end{align}

\begin{align}
    \begin{autobreak}
        \rho_7^m = 
        \frac{31 m_s s^5}{45875200 \pi ^7}
        -\frac{13 \qq{s}  s^4}{491520 \pi ^5}
        -\frac{13 \qq{u}  s^4}{245760 \pi ^5}
        -\frac{7 \qgq{s}  s^3}{40960 \pi ^5}
        -\frac{7 \qgq{u}  s^3}{20480 \pi ^5}
        +\frac{29 \left\langle g_s^2 G^2 \right\rangle  m_s s^3}{2621440 \pi ^7}
        -\frac{5 \left\langle g_s^2 G^2 \right\rangle  \qq{s}  s^2}{24576 \pi ^5}
        -\frac{5 \left\langle g_s^2 G^2 \right\rangle  \qq{u}  s^2}{12288 \pi ^5}
        +\frac{g_s^2 \qq{d} ^2 m_s s^2}{13824 \pi ^5}
        -\frac{\qq{s} ^2 m_s s^2}{256 \pi ^3}
        +\frac{g_s^2 \qq{s} ^2 m_s s^2}{27648 \pi ^5}
        +\frac{9 \qq{u} ^2 m_s s^2}{512 \pi ^3}
        +\frac{g_s^2 \qq{u} ^2 m_s s^2}{6912 \pi ^5}
        +\frac{\qq{d}  \qq{s}  m_s s^2}{64 \pi ^3}
        +\frac{\qq{d}  \qq{u}  m_s s^2}{64 \pi ^3}
        -\frac{\qq{s}  \qq{u}  m_s s^2}{64 \pi ^3}
        -\frac{11 g_s^2 \qq{s} ^3 s}{20736 \pi ^3}
        -\frac{11 g_s^2 \qq{u} ^3 s}{10368 \pi ^3}
        -\frac{11 \qq{d}  \qq{s} ^2 s}{192 \pi }
        -\frac{9 \qq{s}  \qq{u} ^2 s}{64 \pi }
        -\frac{11 g_s^2 \qq{s}  \qq{u} ^2 s}{10368 \pi ^3}
        -\frac{11 g_s^2 \qq{d} ^2 \qq{s}  s}{20736 \pi ^3}
        -\frac{41 \left\langle g_s^2 G^2 \right\rangle  \qgq{s}  s}{196608 \pi ^5}
        -\frac{11 g_s^2 \qq{d} ^2 \qq{u}  s}{10368 \pi ^3}
        -\frac{11 g_s^2 \qq{s} ^2 \qq{u}  s}{5184 \pi ^3}
        -\frac{11 \qq{d}  \qq{s}  \qq{u}  s}{96 \pi }
        -\frac{41 \left\langle g_s^2 G^2 \right\rangle  \qgq{u}  s}{98304 \pi ^5}
        +\frac{11 \qgq{d}  \qq{s}  m_s s}{512 \pi ^3}
        +\frac{77 \qq{d}  \qgq{s}  m_s s}{3072 \pi ^3}
        -\frac{11 \qq{s}  \qgq{s}  m_s s}{1024 \pi ^3}
        +\frac{11 \qgq{d}  \qq{u}  m_s s}{512 \pi ^3}
        -\frac{11 \qgq{s}  \qq{u}  m_s s}{768 \pi ^3}
        +\frac{11 \qq{d}  \qgq{u}  m_s s}{384 \pi ^3}
        -\frac{55 \qq{s}  \qgq{u}  m_s s}{2048 \pi ^3}
        +\frac{27 \qq{u}  \qgq{u}  m_s s}{512 \pi ^3}
        -\frac{3 \qgq{d}  \qq{s} ^2}{128 \pi }
        -\frac{9 \qgq{s}  \qq{u} ^2}{128 \pi }
        -\frac{7 g_s^2 \qgq{s}  \qq{u} ^2}{13824 \pi ^3}
        -\frac{\left\langle g_s^2 G^2 \right\rangle ^2 \qq{s} }{131072 \pi ^5}
        -\frac{g_s^2 \qq{d} ^2 \qgq{s} }{3456 \pi ^3}
        -\frac{g_s^2 \qq{s} ^2 \qgq{s} }{3456 \pi ^3}
        -\frac{7 \qq{d}  \qq{s}  \qgq{s} }{128 \pi }
        -\frac{\left\langle g_s^2 G^2 \right\rangle ^2 \qq{u} }{65536 \pi ^5}
        -\frac{3 \qgq{d}  \qq{s}  \qq{u} }{64 \pi }
        -\frac{3 \qq{d}  \qgq{s}  \qq{u} }{64 \pi }
        -\frac{g_s^2 \qq{d} ^2 \qgq{u} }{1728 \pi ^3}
        -\frac{5 g_s^2 \qq{s} ^2 \qgq{u} }{4608 \pi ^3}
        -\frac{7 g_s^2 \qq{u} ^2 \qgq{u} }{13824 \pi ^3}
        -\frac{\qq{d}  \qq{s}  \qgq{u} }{16 \pi }
        -\frac{9 \qq{s}  \qq{u}  \qgq{u} }{64 \pi }
        -\frac{g_s^3 \left\langle g_s^2 G^2 \right\rangle  \qq{d} ^2 m_s}{73728 \pi ^5}
        +\frac{\left\langle g_s^2 G^2 \right\rangle  \qq{s} ^2 m_s}{2048 \pi ^3}
        -\frac{g_s^3 \left\langle g_s^2 G^2 \right\rangle  \qq{s} ^2 m_s}{49152 \pi ^5}
        +\frac{5 \qgq{s} ^2 m_s}{2048 \pi ^3}
        +\frac{3 \left\langle g_s^2 G^2 \right\rangle  \qq{u} ^2 m_s}{256 \pi ^3}
        +\frac{g_s^3 \left\langle g_s^2 G^2 \right\rangle  \qq{u} ^2 m_s}{18432 \pi ^5}
        +\frac{27 \qgq{u} ^2 m_s}{2048 \pi ^3}
        -\frac{\left\langle g_s^2 G^2 \right\rangle  \qq{d}  \qq{s}  m_s}{1024 \pi ^3}
        +\frac{21 \qgq{d}  \qgq{s}  m_s}{2048 \pi ^3}
        -\frac{\left\langle g_s^2 G^2 \right\rangle  \qq{d}  \qq{u}  m_s}{1024 \pi ^3}
        -\frac{5 \left\langle g_s^2 G^2 \right\rangle  \qq{s}  \qq{u}  m_s}{2048 \pi ^3}
        +\frac{3 \qgq{d}  \qgq{u}  m_s}{256 \pi ^3}
        -\frac{3 \qgq{s}  \qgq{u}  m_s}{1024 \pi ^3}
        +\, .
    \end{autobreak}
\end{align}

\end{document}